# Dissipation and the Relaxation to Equilibrium

Denis J. Evans,[1] Debra J. Searles[2] and Stephen R. Williams[1]

[1] Research School of Chemistry, Australian National University, Canberra, ACT 0200, Australia

[2] Queensland Micro- and Nanotechnology Centre and School of Biomolecular and Physical Sciences, Griffith University, Brisbane, Qld 4111 Australia

**Abstract**

Using the recently derived Dissipation Theorem and a corollary of the Transient Fluctuation Theorem (TFT), namely the Second Law Inequality, we derive the unique time independent, equilibrium phase space distribution function for an ergodic Hamiltonian system in contact with a remote heat bath. We prove under very general conditions that any deviation from this equilibrium distribution breaks the time independence of the distribution. Provided temporal correlations decay, and the system is ergodic, we show that *any* nonequilibrium distribution that is an even function of the momenta, eventually relaxes (not necessarily monotonically) to the equilibrium distribution. Finally we prove that the negative logarithm of the microscopic partition function is equal to the thermodynamic Helmholtz free energy divided by the thermodynamic temperature and Boltzmann's constant. Our results complement and extend the findings of modern ergodic theory and show the importance of dissipation in the process of relaxation towards equilibrium.

The foundations of statistical mechanics are still not completely satisfactory. Textbook derivations of the canonical phase space distribution functions lag a long way behind modern ergodic theory but their derivations fall into two basic categories. The first approach [1-3] is to propose a microscopic definition for the entropy and then to show that the standard canonical distribution function can be obtained by maximising the entropy subject to the constraints that the distribution function should be normalized and that the average energy is constant. The choice of the second constraint is completely subjective due to the fact that at equilibrium, the average of any phase function is fixed. This "derivation" is therefore flawed.

The second approach begins with Boltzmann's postulate of equal a priori probability in phase space for the microcanonical ensemble [2-5] and then derives an expression for the most probable distribution of states in a small subsystem within a much larger microcanonical system. A variation on this approach is to simply postulate a microscopic expression for the Helmholtz free energy [3] via the partition function.

The relaxation of systems to equilibrium is also fraught with difficulties. The first reasonably general approach to this problem is summarized in the Boltzmann H-theorem. Beginning with the definition of the H-function, Boltzmann proved that the Boltzmann equation for the time evolution of the single particle probability density in an ideal gas, implies a monotonic decrease in the H-function [2, 4, 6]. There are at least two problems with this. Firstly the Boltzmann equation is only valid for an ideal gas. Secondly and more problematically, unlike Newton's equations the Boltzmann equation itself is not time reversal symmetric. Some recent work on relaxation to equilibrium has been in the context of relating large deviations to relaxation phenomena and some links with fluctuation theorems have been discussed [7].

The early 1930's saw significant progress in ergodic theory with the proof by Birkoff of the Individual Ergodic Theorem (see [8] for a thorough discussion). Then in 1968 Arnold and Avez [9] showed that if dynamical systems are mixing then correlation functions of two zero mean variables vanish at long times, and furthermore if a dynamical system is ergodic and mixing then in the long time limit the system relaxes to a unique equilibrium state. In 1970 Sinai proved ergodicity for a number of different hard disc and hard sphere systems [8]. He also proved that for these systems the equilibrium state is the usual microcanonical distribution. However it must be said that the ergodic theory of dynamical systems is highly technical and thus these results have largely been restricted to the mathematical physics community. To this day very few dynamical systems have been proved to be mixing and ergodic.

In the present paper we employ a number of new relatively non-technical but exact results that provide new insight into the relaxation to equilibrium. The Transient Fluctuation Theorem (TFT) of Evans and Searles [10] has been derived and its predictions confirmed in laboratory experiments [11]. It is valid for arbitrary densities. Most importantly it can be derived using time reversible microscopic dynamics. The TFT is closely related to a number of other relations [12-14] however the TFT is the only one suited for studying relaxation to equilibrium from a nonequilibrium ensemble.

The TFT is remarkable in that it represents one of the few exact results that apply to nonequilibrium systems far from equilibrium. It provides a generalized form of the 2nd Law of Thermodynamics that applies to small systems observed for short periods of time [10]. The TFT gave the first rigorous explanation of how irreversible macroscopic behaviour arises from time reversible deterministic dynamics and therefore resolves the long-standing Loschmidt paradox [6]. More recently it has been shown that the dissipation function that is the argument of the TFT is also the argument of transient time correlation functions whose time integrals give exact expressions for the linear and nonlinear response [15] of many particle systems.

In the present paper we show how the dissipation function, the TFT and its corollary the Second Law Inequality [16] can be used to resolve certain issues surrounding the relaxation to equilibrium in classical systems. We show that the relaxation of ergodic systems to equilibrium need not be a monotonic process. In spite of this, the relaxation to equilibrium will on average occur in ergodic systems with finite correlation times.

Consider a classical system of N interacting particles in a volume V. The microscopic state of the system is represented by a phase space vector of the coordinates and momenta of all the particles, $\{\mathbf{q}_1, \mathbf{q}_2, ..\mathbf{q}_N, \mathbf{p}_1, ..\mathbf{p}_N\} \equiv (\mathbf{q}, \mathbf{p}) \equiv \Gamma$ where $\mathbf{q}_i, \mathbf{p}_i$ are the position and momentum of particle *i*. Initially (at $t = 0$), the microstates of the system are distributed according to a normalized probability distribution function $f(\Gamma, 0)$. To apply our results to realistic systems, we separate the *N* particle system into a system of interest and a wall region containing $N_W$ particles. Within the wall a subset of $N_{th}$, particles is subject to a fictitious thermostat. The thermostat employs a switch, $S_i$, which controls how many and which particles are thermostatted, $S_i = 0; \quad 1 \leq i \leq (N - N_{th})$, $S_i = 1; (N - N_{th} + 1) \leq i \leq N, N_{th} \leq N_W$. We define the thermostat kinetic energy as

$$K_{th} \equiv \sum_{i=1}^{N} S_i \frac{p_i^2}{2m_i}, \tag{1}$$

and write the equations of motion for the composite N-particle system as

$$\dot{\mathbf{q}}_i = \mathbf{P}_i / m_i$$
$$\dot{\mathbf{p}}_i = \mathbf{F}_i(\mathbf{q}) - S_i(\alpha \mathbf{p}_i + \boldsymbol{\gamma}_{th}) \qquad (2)$$
$$\dot{\alpha} = \left[\frac{2K_{th}}{d_C(N_{th}-1)k_B T_{th}} - 1\right]\frac{1}{\tau^2}$$

where $\mathbf{F}_i(\mathbf{q}) = -\partial \Phi(\mathbf{q})/\partial \mathbf{q}_i$ is the interatomic force on particle i, $\Phi(\mathbf{q})$ is the interparticle potential energy, $-S_i \alpha \mathbf{p}_i$ is a deterministic time reversible Nosé-Hoover thermostat [17] used to add or remove heat from the particles in the reservoir region through introduction of an extra degree of freedom described by $\alpha$, $T_{th}$ is the target parameter that controls the time averaged kinetic energy of the thermostatted particles, $d_C$ is the Cartesian dimension of the system and $\tau$ is the time constant for the Nosé-Hoover thermostat. The vector $\boldsymbol{\gamma}_{th} = \frac{1}{N_{th}}\sum_{i=1}^{N} S_i \mathbf{F}_i$ ensures that the macroscopic momentum of the thermostatted particles is a constant of the motion, which we set to zero. Note that the choice of thermostat is arbitrary, *e.g.* we could use some other choice of thermostat, such as one obtained by use of Gauss' Principle of Least Constraint [17] to fix $K_{th}$, and arrive at essentially the same results. In order to simplify the notation we introduce an extended phase space vector $\boldsymbol{\Gamma}^* \equiv (\boldsymbol{\Gamma}, \alpha)$. In the absence of the thermostatting terms the (Newtonian) equations of motion preserve the phase space volume, $\Lambda \equiv (\partial/\partial \boldsymbol{\Gamma}^*) \cdot \dot{\boldsymbol{\Gamma}}^* = 0$: a condition known as the adiabatic incompressibility of phase space, or AI$\Gamma$ [14]. The equations of motion for the particles in the system of interest are quite natural. The equations of motion for the thermostatted particles are supplemented with unnatural thermostat and force terms. Equations (2) are time reversible and heat can be either absorbed or given out by the thermostat. Similar constructions have been applied in various studies (see, for example, [14, 18]). Of course, if $S_i = 1$ for all *i*, we obtain a homogeneously thermostatted system that is often studied [15].

The derivation of the Evans-Searles TFT [10] considers the response of a system that is initially described by some phase space distribution. The initial distribution is not necessarily at equilibrium however we assume it is an even function of the momenta. We write the initial distribution as

$$f(\mathbf{\Gamma}^*, 0) = \frac{\exp[-F(\mathbf{\Gamma}^*)]}{\int d\mathbf{\Gamma}^* \exp[-F(\mathbf{\Gamma}^*)]}, \tag{3}$$

where $F(\mathbf{\Gamma}^*)$ is a single valued real function which is even in the momenta. The TFT states that provided the system satisfies the condition of ergodic consistency [10], the time averaged dissipation function $\bar{\Omega}_t(\mathbf{\Gamma}^*(0))$, defined as

$$\int_0^t ds\, \Omega(\mathbf{\Gamma}^*(s)) \equiv \ln\left(\frac{f(\mathbf{\Gamma}^*(0),0)}{f(\mathbf{\Gamma}^*(t),0)}\right) - \int_0^t ds\, \Lambda(\mathbf{\Gamma}^*(s))$$
$$\equiv \bar{\Omega}_t(\mathbf{\Gamma}^*(0))t \tag{4}$$

satisfies the following time reversal symmetry [10]:

$$\frac{p(\bar{\Omega}_t = A)}{p(\bar{\Omega}_t = -A)} = \exp[At]. \tag{5}$$

The derivation of the TFT using the exact Liouville equation is straightforward and has been reviewed a number of times [10].

It is important to remember that the existence of the dissipation function $\Omega(\mathbf{\Gamma}^*(0))$ at a phase point $\mathbf{\Gamma}^*(0)$, requires that $f(\mathbf{\Gamma}^*(0),0) \neq 0$. The existence of the integrated form of the dissipation function requires that the dynamics is ergodically consistent (*i.e.* $f(\mathbf{\Gamma}^*(t),0) \neq 0$ for all $\mathbf{\Gamma}^*(0)$ and $t$ for which $f(\mathbf{\Gamma}^*(0),0) \neq 0$). There are systems that fail to satisfy this condition [15]. The existence of the dissipation function (4) only requires that the initial distribution is normalizable and that ergodic consistency holds. To prove (5) requires two additional conditions: the dynamics must be time reversal symmetric and the initial distribution function must be an even function of the momenta.

The TFT leads to a number of corollaries such as the Second Law Inequality [16],

$$\left\langle \overline{\Omega}_t \right\rangle_{f(\mathbf{\Gamma},0)} \geq 0, \quad \forall\, t, f(\mathbf{\Gamma},0), \tag{6}$$

and the NonEquilibrium Partition Identity [19], *i.e.* $\left\langle e^{-\overline{\Omega}_t t} \right\rangle_{f(\mathbf{\Gamma},0)} = 1, \quad \forall\, t > 0$. The notation $\left\langle \ldots \right\rangle_{f(\mathbf{\Gamma},0)}$ implies that the ensemble average is taken over the ensemble defined by the initial distribution $f(\mathbf{\Gamma},0)$, Eq. (3). For simplicity we have dropped the asterisk from the phase space vector in (6) and all that follows.

We have recently derived the Dissipation Theorem [15], a generalization of response theory to handle arbitrary initial distributions, which shows that, as well as being the subject of the TFT, the dissipation function is also the central argument of both linear (*i.e.* Green-Kubo theory) and nonlinear response theory. The Dissipation Theorem gives the following exact expression for the time dependent N-particle distribution function:

$$f(\mathbf{\Gamma}(0),t) = \exp[-\int_0^{-t} ds\, \Omega(\mathbf{\Gamma}(s))] f(\mathbf{\Gamma}(0),0) \tag{7}$$

This expression is valid for systems that are driven away from equilibrium by an external field or, as in this paper, systems that have no externally applied field but that may initially have nonequilibrium distributions. An important observation arises from this expression. The necessary and sufficient condition for a distribution function to be time independent is that:

- for ergodic systems [20], the dissipation function must be zero everywhere in phase space and,

- for nonergodic systems, the dissipation function must be zero everywhere within the ergodic *subdomain* in which a given sample system is trapped.

In reference [15] (also see [19, 21]) we proved that we can write averages of arbitrary phase functions as

$$\langle B(t) \rangle_{f(\Gamma,0)} = \langle B(0) \rangle_{f(\Gamma,0)} + \int_0^t ds \langle \Omega(0) B(s) \rangle_{f(\Gamma,0)}. \tag{8}$$

The derivation of (7),(8) from (4) is called the Dissipation Theorem. This Theorem is extremely general. Like the TFT it is valid arbitrarily far from equilibrium. As in the derivation of the TFT the only unphysical terms in the derivation are the thermostatting terms within the wall region. However, because these thermostatting particles can be moved arbitrarily far from the system of interest, the precise mathematical details of the thermostat are unimportant.

For the Nosé-Hoover dynamics (2), consider the initial distribution

$$f(\Gamma,0) \equiv f_C(\Gamma) = \frac{\delta(\mathbf{p}_{th}) \exp[-\beta_{th} H_E(\Gamma)]}{\int d\Gamma\, \delta(\mathbf{p}_{th}) \exp[-\beta_{th} H_E(\Gamma)]} \tag{9}$$

where $H_0(\Gamma)$ is the internal energy of the system, $H_E(\Gamma) = H_0(\Gamma) + \tfrac{1}{2} d_C (N_{th} - 1) k_B T_{th} \alpha^2 \tau^2$ is the so called extended Nosé-Hoover Hamiltonian, $k_B T_{th} \equiv \beta_{th}^{-1}$, and

$\delta(\mathbf{p}_{th}) \equiv \delta(\sum S_i p_{xi})\delta(\sum S_i p_{yi})...\delta(\sum S_i p_{d_ci})$ fixes the total momenta of the thermostatted particles in each Cartesian dimension at zero. We shall call this the canonical distribution even though it includes extra degrees of freedom for the thermostat multiplier $\alpha$.

It is trivial to show that for this distribution and the dynamics (2) the dissipation function, $\Omega_C(\Gamma)$, is identically zero

$$\Omega_C(\Gamma) = 0, \quad \forall \, \Gamma \tag{10}$$

and from (7) we see that this initial distribution is preserved

$$f(\Gamma, t) = f_C(\Gamma), \quad \forall \, \Gamma, t. \tag{11}$$

For ergodic systems [22] we call distributions that are time independent and dissipationless, *equilibrium* distributions. From (4), it can be shown that $\exp(\bar{\Omega}_t(\Gamma(0))t)$ is equal to the ratio of probability of seeing an infinitesimal set of phase space trajectories centred on $\Gamma(0)$ compared to the probability of observing (in the same distribution) the conjugate set of anti-trajectories centred on $M^T \Gamma(t)$ where $M^T(\mathbf{q}, \mathbf{p}) \equiv (\mathbf{q}, -\mathbf{p})$ is the time reversal map of the phase point [19]. Therefore the dissipation can only be zero everywhere in the phase space when these probabilities are equal (*i.e.* in an equilibrium system). By definition, in an equilibrium system you cannot distinguish the direction of time [23].

Now consider an arbitrary deviation from the canonical distribution

$$f(\Gamma, 0) \equiv \frac{\delta(\mathbf{p}_{th})\exp[-\beta_{th} H_E(\Gamma) - \gamma g(\Gamma)]}{\int d\Gamma \, \delta(\mathbf{p}_{th})\exp[-\beta_{th} H_E(\Gamma) - \gamma g(\Gamma)]} \tag{12}$$

where $g(\Gamma)$ is an arbitrary integrable real function that, since $f(\Gamma,0)$ must be an even function of the momenta, is also even in the momenta. Without loss of generality we assume $0 \leq \gamma$.

For such a system evolving under our dynamics (2), the time integrated dissipation function is

$$\bar{\Omega}_t(\Gamma(0))t = \gamma[g(\Gamma(t)) - g(\Gamma(0))] \equiv \gamma \Delta g(\Gamma(0),t) \tag{13}$$

and (7) becomes

$$f(\Gamma(0),t) = \exp[-\gamma \Delta g(\Gamma(0),-t)] f(\Gamma(0),0). \tag{14}$$

Thus, if $g$ is not a constant of the motion there is dissipation and the distribution function cannot be an equilibrium distribution. If the system is ergodic the only possible unchanging distribution function will be that where $g(\Gamma(t)) = 0, \forall \Gamma(0), t$. If there are constants of the motion that are not factored into the distribution (9), the system cannot be ergodic since the phase space can, be broken down into a number of phase space subdomains characterized by different fixed values of $\gamma g(t)$.

Summarising: if the system is ergodic there is a single unique time symmetric, equilibrium state characterized by being dissipationless everywhere in the phase space(10). For the system considered here that distribution is the canonical distribution (9). Thus we have derived an expression for the unique equilibrium state corresponding to the thermostatted equations of motion and shown that it takes on the standard form for the canonical distribution, modulo the facts that: in the thermostatting region the momentum is a constant of the motion that is set to zero, and that there is an extended degree of freedom for the thermostat.

This completes our first-principles derivation of the equilibrium distribution function. We now consider the question of relaxation towards equilibrium.

The dissipation function satisfies the Second Law Inequality (6),

$$\gamma \langle \Delta g(\mathbf{\Gamma}(0),t) \rangle = \int_0^\infty dA \ A(1-e^{-A}) p[\gamma \Delta g(\mathbf{\Gamma}(0),t) = A]$$

$$\geq 0$$

(15)

which can only take on a value of zero when $\gamma \Delta g(\mathbf{\Gamma}(0),t) = 0, \forall \mathbf{\Gamma}(0)$. The proof that $\Delta g(\mathbf{\Gamma}(0),t)$ must be zero *everywhere* in phase space is obvious from the first line of (15) (*i.e.* if $p[\gamma \Delta g(\mathbf{\Gamma}(0),t) = A]$ is non-zero for any $A > 0$, then $p[\gamma \Delta g(\mathbf{\Gamma}(0),t) = A]A(1-e^{-A}) > 0$ and the integrand as well as the integral will be positive). Thus in an ergodic system if the initial distribution differs in any way from the canonical distribution there will be dissipation and on *average* this dissipation is *positive*. This remarkable result is true for arbitrary $\gamma, g(\mathbf{\Gamma})$.

We assume the system is ergodic and apply the Dissipation Theorem. Substituting (13) into (8) gives:

$$\langle \Delta g(\mathbf{\Gamma}(0),t) \rangle_{f(\mathbf{\Gamma}(0),0)} = \gamma \int_0^t ds \ \langle \dot{g}(0) g(s) \rangle_{f(\mathbf{\Gamma},0)} \geq 0$$

(16)

where the ensemble averages are taken with respect to the distribution function (12).

We assume that at sufficiently long time, $t_c$ temporal correlations have decayed to zero, so that for $t > t_c$, we can write :

$$\langle \Delta g(\boldsymbol{\Gamma}(0),t)\rangle_{f(\boldsymbol{\Gamma},0)} = \gamma \int_0^{t_c} ds \, \langle \dot{g}(0)g(s)\rangle_{f(\boldsymbol{\Gamma},0)} + \gamma \int_{t_c}^{t} ds \, \langle \dot{g}(0)\rangle_{f(\boldsymbol{\Gamma},0)} \langle g(s)\rangle_{f(\boldsymbol{\Gamma},0)}$$

$$= \gamma \int_0^{t_c} ds \, \langle \dot{g}(0)g(s)\rangle_{f(\boldsymbol{\Gamma},0)} = \langle \Delta g(\boldsymbol{\Gamma}(0),t_c)\rangle_{f(\boldsymbol{\Gamma},0)} \quad (17)$$

$$\geq 0$$

The last term on the first line is zero since $g(\boldsymbol{\Gamma})$ is an even function of the momenta and therefore $\langle \dot{g}(0)\rangle_{f(\boldsymbol{\Gamma},0)} = 0$. This means that $\langle \dot{g}(t)\rangle_{f(\boldsymbol{\Gamma},0)} = 0, t > t_c$ and $\langle g(t)\rangle, t > t_c$ does not change with time, *i.e.* $\langle g(t)\rangle_{f(\boldsymbol{\Gamma},0)} = const, t > t_c$. If we take $t_c$ to be a new time origin then the system has no average dissipation after this time, and as discussed after equation (15) the system must be in equilibrium.

From (12) we see that if the perturbation relaxes conformally in phase space (the deviation function simply scales by a single time dependent parameter $\gamma(t)$), we see that in order to increase the value of $\langle g(t)\rangle_{f(\boldsymbol{\Gamma},0)}$ we must decrease the magnitude of the scale parameter $\gamma$, implying that the system is closer to the equilibrium distribution. This also means that we can continuously redefine a new time origin. In this case equation (17) implies there is a monotonic relaxation to equilibrium.

We have seen above for ergodic systems the dissipationless distribution is the unique equilibrium state. Because the average integrated dissipation is positive, even in the non-conformal case the system must initially move towards equilibrium; in fact from (16) we can prove that

$$\lim_{t \to 0^+} \langle \dot{g}(t)\rangle_{f(\boldsymbol{\Gamma},0)} = \gamma \langle \dot{g}^2(0)\rangle_{f(\boldsymbol{\Gamma},0)} \geq 0. \quad (18)$$

This proves that initially, on average, the system must move towards, rather than away, from equilibrium. At later times the system may move, for a short time, away from equilibrium (*e.g.*

as in the case of an under-damped oscillator) but such movement is never enough to make the time integrated average dissipation negative. The time integrated average dissipation from the initial state to any intermediate state (including the final equilibrium state) is non-negative. At any sufficiently later instant in the relaxation process, the *instantaneous* dissipation may be negative. This shows that, in general, the relaxation process may not be monotonic in time. Such non-monotonic relaxation is extremely common in Nature.

If the system is not ergodic then by definition the system relaxes to a number of possible quasi-equilibrium states. Such states break up phase space into ergodic subdomains within which the relative probabilities are canonically distributed but between which, the relative weights are history dependent non-Boltzmann distributed [24].

This completes our first-principles derivation of the relaxation towards equilibrium. We call this proof, the Relaxation Theorem. The theorem was proved assuming, time reversible microscopic dynamics, ergodicity, ergodic consistency, the initial distribution is even under a time reversal mapping and the decay of correlations. This latter condition may be relaxed to merely require the convergence of the transient time correlation integrals in (16). In this case equilibrium will be approached in the limit of infinite time. If the correlation integral in (16) diverges, the system cannot be expected to ever relax to equilibrium.

It is reassuring that the conditions for the Relaxation Theorem to apply are effectively the same conditions as are known from ergodic theory to lead to the relaxation to equilibrium states [8]. However, the result is obtained using quite a different approach. In addition we have learned more about the relaxation process than ergodic theory approach has taught us so far.

In summary we have demonstrated that for any ergodic Hamiltonian system of fixed volume and fixed number of particles, in contact with a heat reservoir whose initial (nonequilibrium) distribution is even under time reversal symmetry:

- there is a *unique* dissipationless state, and this state has the canonical distribution, (9) [Although a Nosé-Hoover thermostat was used in this derivation, essentially the same result is obtained with other thermostatting mechanisms such as a Gaussian isokinetic thermostat];

- in ergodic systems with finite temporal correlations the system eventually relaxes to equilibrium;

- this relaxation towards equilibrium is *not* necessarily monotonic [We note that the Boltzmann H-theorem that applies to dilute gases only, implies a monotonic relaxation to equilibrium, thus the Relaxation Theorem allows for much more complex behaviour as seen experimentally.] and,

- inequality, Eq. (16), for the time integrated average dissipation shows that $\langle \Delta g(\Gamma(0),t) \rangle_{f(\Gamma(0),0)} \geq 0$, and if the deviation function relaxes conformally, the system will relax to equilibrium monotonically;

- equality (18) shows that the initial ensemble average response is always towards, rather than away from, equilibrium.

We have also shown quite generally that for ergodic dynamical systems obeying time reversible dynamics, states have properties that are time reversal symmetric (*i.e.* probabilities of observing any set of trajectories and its conjugate set of antitrajectories are equal) if and only if the dissipation function is zero everywhere in phase space. If there is dissipation anywhere in the phase space the distribution function is time dependent and the system cannot be in equilibrium.

Our last task is to connect our equilibrium distribution function with thermodynamics. Some textbooks [4] use circular arguments to prove the connection between thermodynamic and statistical mechanical variables while most textbooks [5], apply a sort of pattern recognition to achieve this purpose. Here we give a simple and direct derivation.

We *postulate* that,

$$A(T = T_{th}, N, V) = Q(T_{th}, N, V)$$
$$\equiv -k_B T_{th} \ln\left[\int d\Gamma\, \delta(\mathbf{p}_{th}) \exp[-\beta_{th} H_E(\Gamma)]\right] \quad (19)$$

That is when $T_{th} = T$ the Helmholtz free energy A(T), at the thermodynamic temperature T, is equal to the value of the statistical mechanical expression $Q(T_{th})$ that is defined in (19). From thermodynamics we note that the Helmholtz free energy satisfies the differential equation

$$U = A - T\frac{\partial A}{\partial T} \quad (20)$$

where U is the internal energy. Whereas if we differentiate Q with respect to $T_{th}$ we see that

$$\langle H_E \rangle = Q - T_{th}\frac{\partial Q}{\partial T_{th}} \quad (21)$$

Since $U = \langle H_E \rangle$ (internal energy is an entirely mechanical quantity as the First Law shows) and noting that when $T = T_{th} = 0$, that $A(0) = U(0) = Q(0)$, we observe, treating $T, T_{th}$ as integration parameters x, that A and Q satisfy the same differential equations with the same initial $x = 0$ condition and therefore $A(T) = Q(T_{th})$ and our hypothesis (19) is proved [25, 26].

We have shown that the Evans-Searles Fluctuation Theorem, the Second Law Inequality and the Dissipation Theorem provide useful insights into the relaxation to equilibrium. These relations can be combined to prove the Relaxation Theorem that shows, subject to ergodicity and the decay of correlations, that nonequilibrium distributions that are even functions of the momenta relax, on average to equilibrium. Although these conditions are essentially the same as those used in ergodic theory to predict the relaxation to equilibrium, our results explicitly admit the possibility that the relaxation to equilibrium, in accord with common experience, may not proceed monotonically. Our theory also shows that initially, a nonequilibrium system *must* on average move towards, rather than away from, equilibrium. Our results show that the dissipation function is a key function in both defining the nature of equilibrium and in describing, via transient time correlation functions, the process of relaxation to equilibrium. That same function is also the argument of both the Evans-Searles Fluctuation Theorem and the Dissipation Theorem.

**Acknowledgement**

The authors would like to thank Lamberto Rondoni for comments on the status of ergodic theory.